\documentclass{chi2009}
\usepackage{times}
\usepackage{graphicx}
\usepackage{hyperref}
\graphicspath{{pics/}}
\usepackage{url}
\usepackage{array}
\usepackage{fancyhdr}
\usepackage{color}

\newcommand{\capspace}{-1em}	

\begin{document}

%
\toappear{FXPAL-TR-10-003. Copyright (2010) held by the authors.}

\title{mVideoCast: Mobile, real time ROI detection and streaming}
%
%
%
%
%

\numberofauthors{1} 
%
\author{
%
%
\alignauthor
Scott Carter, Laurent Denoue, John Adcock\\
       \affaddr{FX Palo Alto Laboratory}\\
       \affaddr{3400 Hillview Ave.}\\
       \affaddr{Palo Alto, CA 94304 USA}\\
       \email{(carter,denoue,adcock)@fxpal.com}
}

\date{11 April 2010}

\maketitle

\begin{abstract}
A variety of applications are emerging to support streaming video from mobile devices. However, many tasks can benefit from streaming specific content rather than the full video feed which may include irrelevant, private, or distracting content. We describe a system that allows users to capture and stream targeted video content captured with a mobile device. The application incorporates a variety of automatic and interactive techniques to identify and segment desired content in the camera view, allowing the user to publish a more focused video.
\end{abstract}




\section{Introduction}

\begin{figure}[t]
\centering
\begin{tabular}{cc} 
 \includegraphics[width=1.37in]{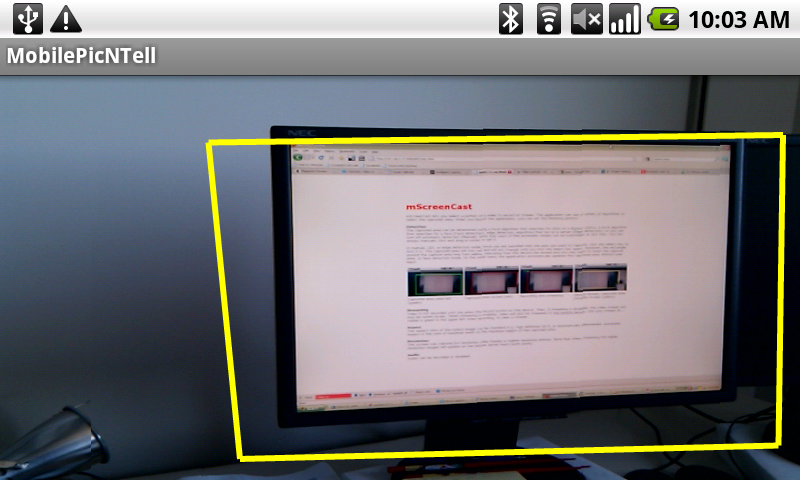} &  
 \includegraphics[width=1.37in]{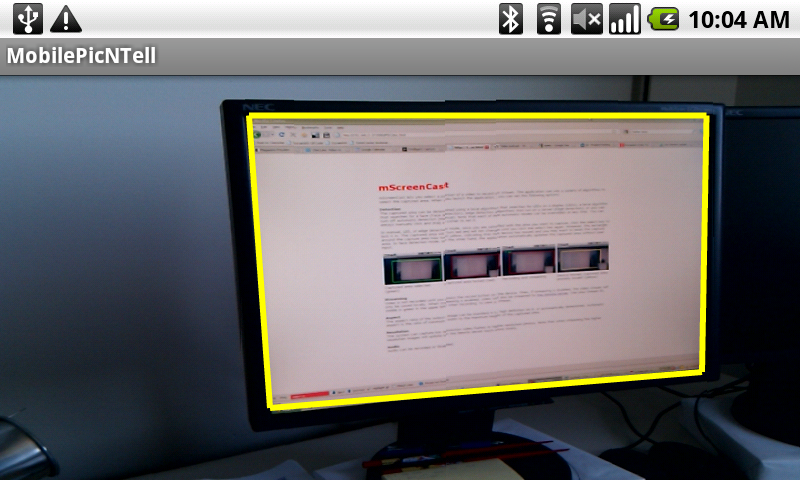} \\  
(a) & (b) \\
 \includegraphics[width=1.37in]{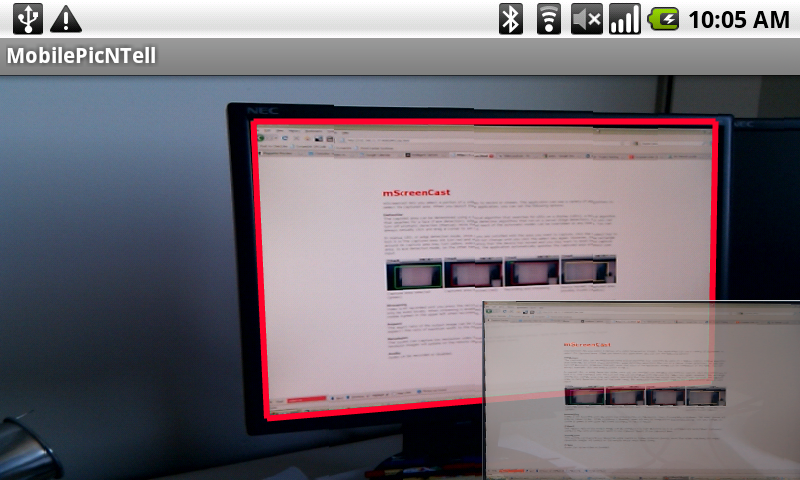} & 
 \includegraphics[width=1.37in]{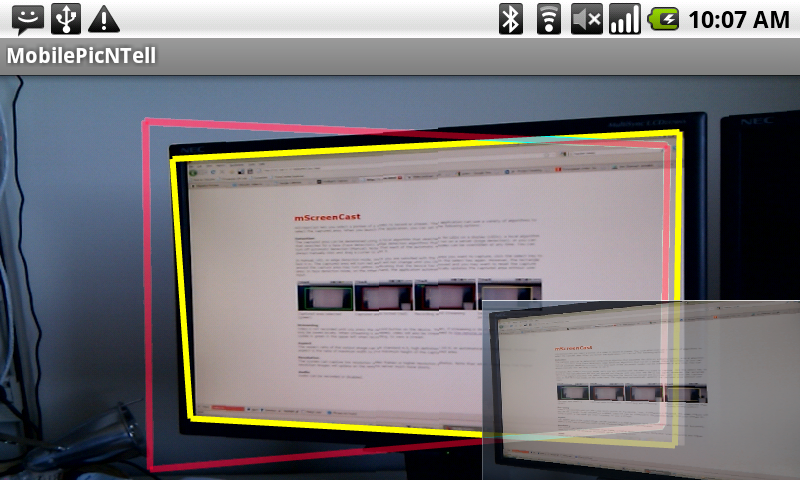} \\
(c) & (d) \\ 
 \includegraphics[width=1.37in]{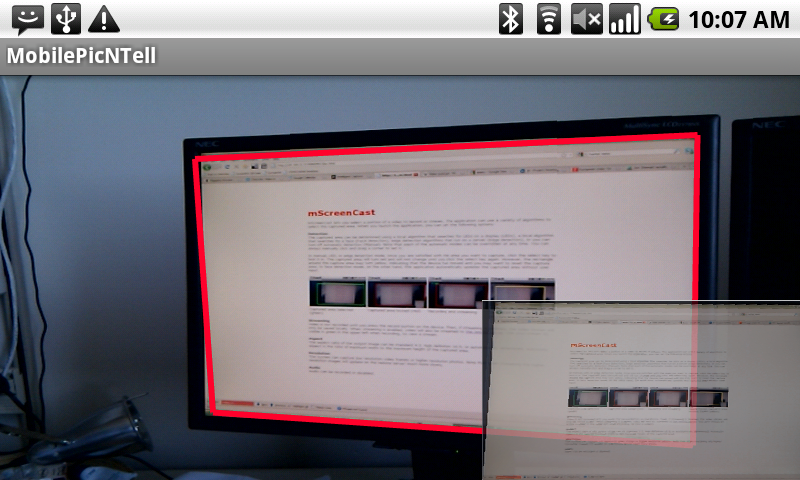} &
 \\
(e)  
\end{tabular}
\vspace{\capspace}
\caption{Screen detection. The application's first estimate shown in yellow (a) is inaccurate, but the second (b) correctly identifies the screen. When the user locks the ROI it changes color to red and the application displays a transparent image of the warped and cropped ROI in the lower-right, (c). As the user repositions the device, the application detects the motion and begins finding a new ROI, (d). The user can press a button to lock the ROI to the new guess, (e). }
\label{fig:screendetect}
\end{figure}

\enlargethispage{\baselineskip}

As the processing power of mobile devices improves, they are being used for more computationally intensive tasks. Services have begun to offer live video streaming from mobile devices (e.g., Qik \cite{Qik}), potentially allowing anyone to stream any event anywhere at anytime. However, as we have seen in the desktop world, unfiltered streaming, while useful, is not appropriate for every task. By filtering streamed content, presenters can better focus the audience's attention, improve bandwidth efficiency, and mitigate privacy concerns. Furthermore, mobile content capture faces challenges not present on desktops -- for example the recording device may be off-axis from the desired content and may be hand-held and therefore unstable.

\pagebreak

The system we present, mVideoCast, helps filter and correct video captured and streamed from a mobile device (see Figure \ref{fig:screendetect}). The application can detect, segment, and stream content shown on screens or boards, faces, or arbitrary regions. This can allow anyone to stream task-specific content without needing to develop hooks into external software (e.g., screen recorder software). While past work has supported streaming ROIs from video, mVideoCast is the first tool to do so live from a mobile device. In this paper we describe the system, algorithms we use to detect regions-of-interest (ROIs) in video frames, and our experiences with early prototypes of the application.

\section{mVideoCast}

mVideoCast is architected as a client-server system in which mobile clients publish content to a remote machine. The mobile application, implemented on the Android platform, can both record captured media on the device locally as well as stream content to a remote server in real time, allowing remote users to view live content. Users can control when the application is recording content locally and when content is streaming live to a remote server.

\subsection{Capturing and correcting the image frame}

\begin{figure}[tb]
\centering
\begin{tabular}{cc} 
 \includegraphics[width=2in]{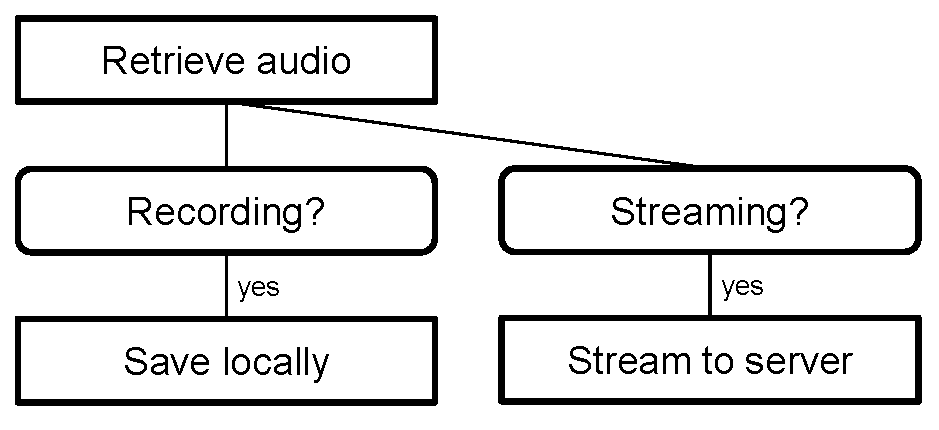} \\
 (a)\\
 \includegraphics[width=3.1in]{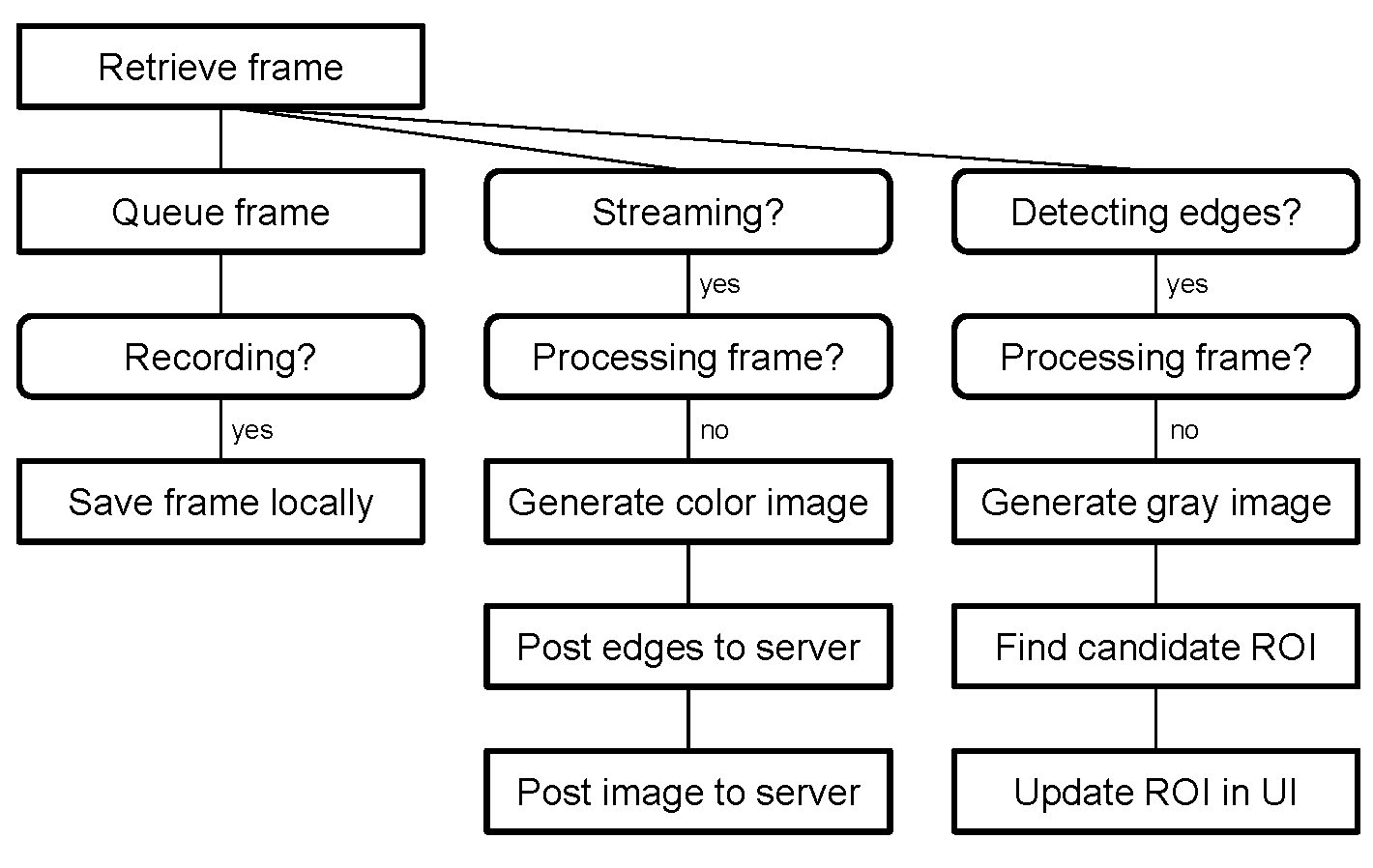} \\
 (b)\\
\end{tabular}
\vspace{\capspace}
\caption{Data flow. Audio (a) can be saved locally and streamed to a remote server. Video frames (b) can be saved, streamed, and used to detect ROIs.}
\label{fig:system}
\end{figure}

The mobile application extracts a quadrilateral ROI from each video frame, warping and cropping the frame to match a pre-defined output resolution and aspect ratio (see Figure \ref{fig:system}). The application runs three separate threads in order to save content locally, stream content to a remote server, and detect ROIs. The application first opens the camera and requests to receive preview frames. It stores received frames in a queue, and if it is in record mode it saves queued images locally. It then pushes frame events to the other two threads. The stream thread checks to see that it is in stream mode and that it has processed the last event. It then generates a compressed color image of the frame and posts it to a remote server (full images are sent, rather than only the ROI, to support potential post hoc edits to the stream). The detection thread similarly checks that detection is enabled and that it has processed the last frame. It then generates a grayscale image and runs the screen detection algorithm (methods for defining the ROI are described in the next section). Once a candidate quadrilateral is found, it is set in the main thread and shown on the display as an overlay drawn over the video preview (see Figure \ref{fig:screendetect}).

\subsection{Audio}
\begin{figure}[tb]
\centering
\begin{tabular}{c} 
 \includegraphics[width=2.5in]{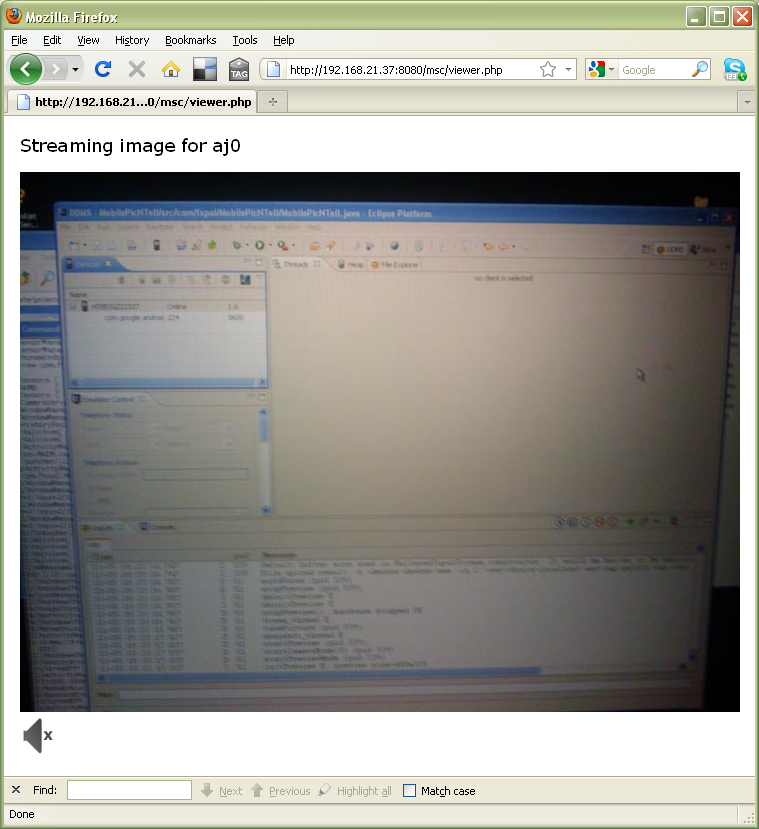} \\
\end{tabular}
\vspace{\capspace}
\caption{Viewing a stream. The server publishes a web page for each client that shows the cropped and warped ROI from the latest frame as well as the client's audio stream (disabled here).}
\label{fig:stream}
\end{figure}

Because video frames undergo several processing steps and may not maintain a constant frame rate, audio is captured, saved, and streamed separately. While recording, captured audio is saved to a local file. While streaming, audio is Speex~\cite{speex} encoded and forwarded to an Icecast~\cite{icecast} server running remotely. A web page associated with each user merges the audio with processed frames (see Figure \ref{fig:stream}).

\pagebreak

\section{Selecting a region}
Users can define a ROI using a variety of methods ranging from fully manual to fully automatic.

\subsection{Manual selection}
Users can set the ROI by tapping on the screen to set the four corners of the quadrilateral. The corner nearest the tapped point is set to the tapped location. There are also shortcuts to set certain standard sizes more quickly; clicking on the upper-left and lower-right corners of the quadrilateral in rapid succession define a rectangle, and double-tapping anywhere on the screen sets the entire preview frame as the captured region.

\subsection{Light tags}
\begin{figure}[tb]
\centering
\begin{tabular}{c} 
 \includegraphics[width=3in]{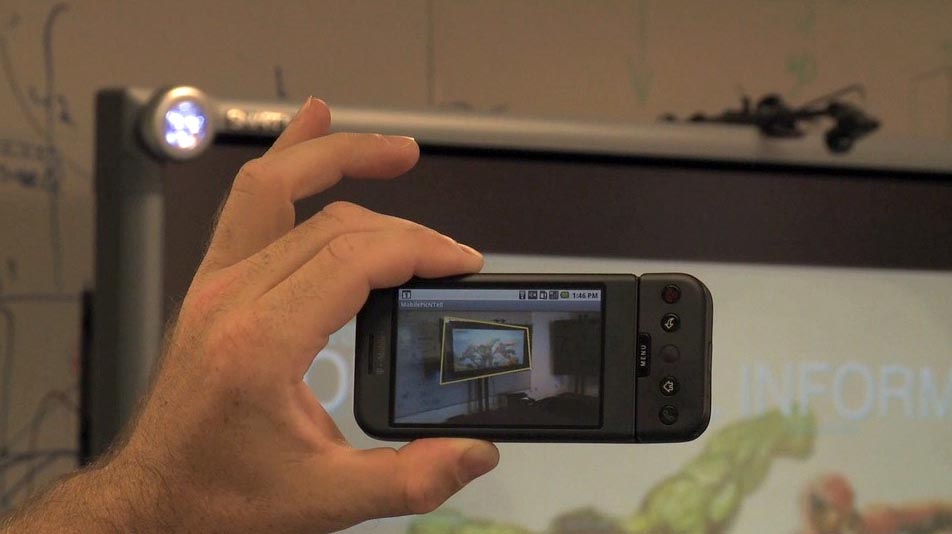} \\
\end{tabular}
\vspace{\capspace}
\caption{Light tag detection. The mobile application can search for light tags (e.g., LEDs) attached to displays.}
\label{fig:LED}
\end{figure}

If the user has control of the display he wants to capture, he can switch the mobile application to a mode that detects light tags (e.g., LEDs) attached to the display. In this mode the mobile application detects the bright points corresponding to the light tags to determine the corners of the ROI (see Figure \ref{fig:LED}).

\subsection{Screen detection}
Users can enable a screen detection mode that will automatically attempt to determine screen regions within frames. We make use of the JJIL toolkit \cite{JJIL} in order to detect screens in a frame as follows: 1) Run a Canny edge detector over a grayscale version of the frame; 2) Remove all but the top 5\% most significant edges from the result; 3) Divide the remaining points into four regions representing the top-half, bottom-half, left-half, and right-half of the image; 4) Send the subregions to a Hough line fit algorithm to find the dominant line in each subregion; 5) Construct a quadrilateral from the resulting lines. 

This approach is typically not immediately accurate, so while in screen detection mode the application presents the current ROI estimate to the user every few seconds (see Figure \ref{fig:screendetect}). When the user is satisfied with a result, he can use a hardware button to ``lock'' the current ROI coordinates in place. When a user locks an ROI, the application displays a thumbnailed view of the warped and cropped region in the lower-right of the screen.

We anticipate that the user will move around while recording, and after moving the ROI will no longer be accurate. To address this issue, the mobile application continuously monitors the mobile device's onboard accelerometer and compass to detect a change in position. When it determines that the device has moved, it begins generating new potential ROIs. While it generates guesses it also maintains the past ROI. When the user decides that a new estimate is a better match, he can click the hardware button to set it as the current ROI.

The user can also adjust the corners of the quadrilateral at any time using the manual controls described above.

\subsection{Face detection}
\begin{figure}[tb]
\centering
\begin{tabular}{c} 
 \includegraphics[width=3in]{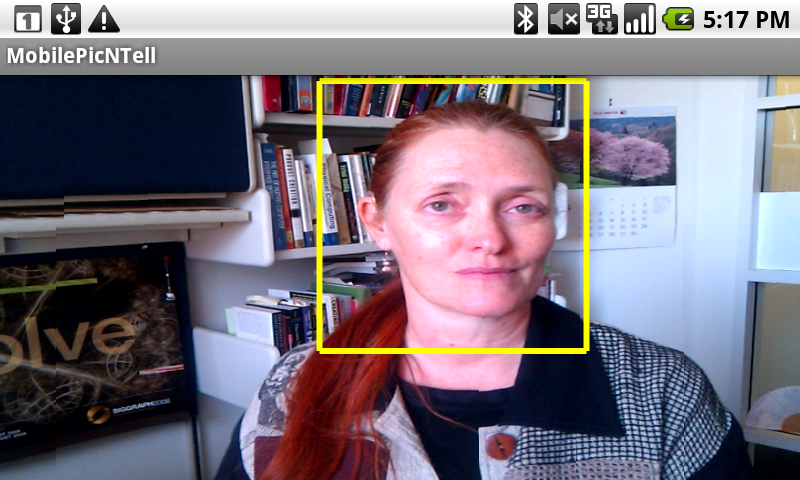} \\
\end{tabular}
\vspace{\capspace}
\caption{Face tracking. The mobile application can set the ROI to be the most salient face in a frame.}
\label{fig:facedetect}
\end{figure}

In another mode the application uses Android's face detector libraries to set the ROI. In this mode, the application automatically updates the ROI with the location of the most salient face detected in each frame (see Figure \ref{fig:facedetect}). 
In this mode the application does not adjust the output aspect ratio and only crops the frame.

\section{Scenarios}

mVideoCast can be used for a variety of tasks.

\begin{description}

\item{\textbf{Reporting:}}
A technology reporter uses his mobile phone's camera to record the screen of a demonstration of new software at a local startup. mVideoCast lets him generate a clean, bandwidth efficient upstream for his followers to see live or asynchronously.

A news reporter is interviewing a subject on the street. By using the face detection mode, mVideoCast allows the reporter to easily stream a focused view of the interviewee, reducing distractors in the video and possibly preserving the privacy of bystanders.

\item{\textbf{Remote demonstrations:}}
A business user wants to present a demonstration of a software application while sitting at a cafe. He starts mVideoCast on his phone and can stream only his screen to the remote participants; the system will crop regions out of the screen such as the cafe surroundings and his croissant.

\item{\textbf{Sharing other mobile screens:}}
A designer wants to document how a web page renders on his iPhone. He uses a second mobile device with mVideoCast to record his iPhone screen. The application picks up the iPhone screen boundaries in the video stream, unwarps it and streams it to remote viewers.

\item{\textbf{Troubleshooting:}}
Alice is trying to send a FAX using the her office's multifunction printer, but the machine stops unexpectedly while processing her paper. She decides to call support and puts her phone on loud-speaker, launches mVideoCast, and points her device at the printer's LCD screen. The software automatically detects the boundaries of the LCD screen, un-warps its image and sends it to a member of the support staff who can more easily view it and guide her to a solution.

\end{description}

\enlargethispage{\baselineskip}
\section{Experience}
We asked four subjects to provide initial feedback on a prototype of the system running on an Android Nexus One device. We asked them to imagine a scenario in which they were sending live video of their desktop monitor to a remote party.
We explained the goal of the system was to help them automatically crop around the ROI, and that manual adjustments were possible by dragging the automatically selected edges to other locations or tapping the screen. 

User feedback provided several potential areas of improvement.  Because many edges are present in the image of a typical computer desktop, the subjects often had to manually readjust the area automatically selected by the system. This proved problematic, as dragging or touching corners with fingers would sometimes hide most of the screen and made people move the device's screen, causing irregularities with the selected ROI. In general, holding the device steady was a challenge, and user's hands would naturally move when they had to look at the object they wanted to capture -- looking at the device's display was not enough. One user would always let the system guess the good tracking rectangles, never locking a guess by pressing the designated button. Another noticed that guesses where sometimes correct, but he did not have time to press the button before the next guess happened. He suggested a ``back'' button to validate previous system guesses. Also, pressing the hardware button resulted in significant device motion, occasionally interfering with the ROI selection. A user suggested replacing this action by a touch in the middle of the detected area.


User feedback also revealed new use cases for the tool. Three subjects had to step far back in order to capture their 30'' monitor, which prompted them to try capturing individual windows within their desktop area. They actually expressed interest in this use case, capturing for example a login window or a web browser window. One user said he would actually like this system to capture an unwarped picture of a window, not necessarily a video recording. Another participant spontaneously and successfully used the application to capture parts of his whiteboard.


\section{Related work}

While other research projects have explored video retargeting, or automatically selecting salient subregions of a video for redisplay on smaller screens such as mobile devices \cite{Liu06}, mVideoCast uniquely allows users to stream specific ROIs from a mobile device. 

There have been a variety of applications that have used ROIs in video in non-mobile contexts. Researchers have investigated user- and group-defined ROIs to control cameras for remote collaboration tasks \cite{Liu02,Song02}. Similarly, the Diver system allows users to create videos from cropped clips of a prerecorded, panoramic video \cite{Pea04}. Other tools have explored automated solutions.
El-Alfy et al. investigated automatically cropping surveillance videos to salient events \cite{Elalfy07}. Another focus of past work is the removal of individuals from video recordings or video conference streams (such as \cite{Chen07}).

There is existing work concerning the detection of documents in still images, including business card detection~\cite{Hua06}. It is becoming more common for compact digital still cameras to include a document mode which will attempt to automatically identify the outline of a document and allow the user to save a perspective-corrected version of the image. The iPhone application, JotNot~\cite{jotnot}, provides a similar quadrilateral region crop-and-correct capability for still images, but requires manual selection of the target corners.

\begin{figure}[tb]
\centering
\begin{tabular}{c} 
 \includegraphics[width=3.1in]{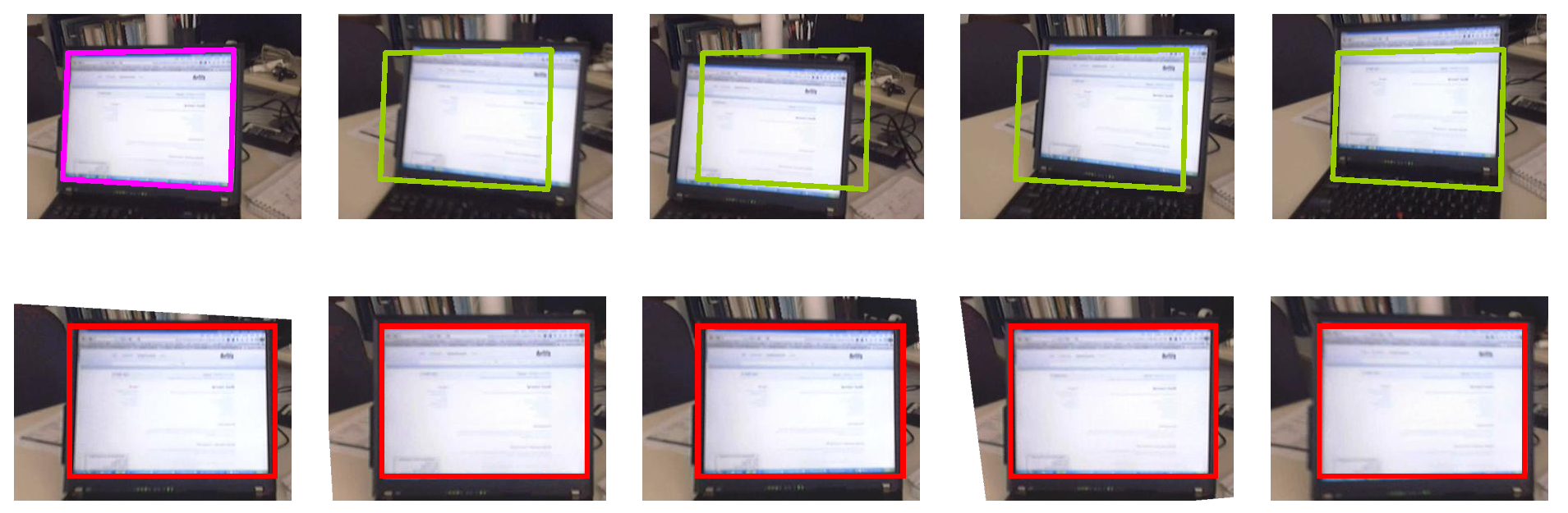} \\
\end{tabular}
\vspace{\capspace}
\caption{Using image stabilization to improve screen detection. We are experimenting with optical flow techniques that can make minor adjustments to the original frames (top) in order to keep the ROI registered without user intervention (bottom).}
\label{fig:flow}
\end{figure}

\section{Conclusion and future work}


In the past remote communication suffered primarily from a lack of bandwidth. Today networked, mobile multimedia devices are ubiquitous, and the core challenge is not how to transmit \emph{more} information but rather how to communicate the \emph{right} information. mVideoCast is a small but important step toward this goal.

We are working to improve mVideoCast in a number of ways. First, we are investigating new single-handed controls for setting the ROI manually. Furthermore, we are exploring the use of optical motion compensation to maintain the alignment of the ROI between explicit detections. The step of locking coordinates provides an anchor to which subsequent frames can be registered. That is, if the ROI is locked at frame 0, subsequent frames 1 ... N can be aligned to this reference frame 0 with well known automatic image registration techniques. For instance, we can compute a transformation given a set of corresponding image coordinates, determined by matching image features (see Figure \ref{fig:flow}). In this way, the initial lock can be used without losing the position due to camera motion.

\bibliographystyle{abbrv}
\bibliography{nudgecam}  

\begin{thebibliography}{10}

\bibitem{Chen07}
D.~Chen, Y.~Chang, R.~Yan, and J.~Yang.
\newblock Tools for protecting the privacy of specific individuals in video.
\newblock {\em EURASIP J. Appl. Signal Process.}, 2007(1):107--107, 2007.

\bibitem{Elalfy07}
H.~El-Alfy, D.~Jacobs, and L.~Davis.
\newblock Multi-scale video cropping.
\newblock In {\em ACM MM '07}, pages 97--106, New York, NY, USA, 2007. ACM.

\bibitem{Hua06}
G.~Hua, Z.~Liu, Z.~Zhang, and Y.~Wu.
\newblock Automatic business card scanning with a camera.
\newblock In {\em International Conference on Image Processing (ICIP)}, Los
  Alamitos, CA, USA, 2006. IEEE.

\bibitem{icecast}
Icecast.org.
\newblock \url{http://www.icecast.org}.

\bibitem{JJIL}
{Jon's Java Imaging Library, for mobile image processing}.
\newblock \url{http://code.google.com/p/jjil/}, 2010.

\bibitem{jotnot}
{J}ot{N}ot.
\newblock \url{http://www.jotnot.com}.

\bibitem{Liu06}
F.~Liu and M.~Gleicher.
\newblock Video retargeting: automating pan and scan.
\newblock In {\em ACM MM '06}, pages 241--250, New York, NY, USA,
  2006. ACM.

\bibitem{Liu02}
Q.~Liu, D.~Kimber, J.~Foote, L.~Wilcox, and J.~Boreczky.
\newblock Flyspec: a multi-user video camera system with hybrid human and
  automatic control.
\newblock In {\em ACM MM '02}, pages 484--492, New York, NY, USA, 2002. ACM.

\bibitem{Pea04}
R.~Pea, M.~Mills, J.~Rosen, K.~Dauber, W.~Effelsberg, and E.~Hoffert.
\newblock The diver project: Interactive digital video repurposing.
\newblock {\em IEEE MultiMedia}, 11(1):54--61, 2004.

\bibitem{Qik}
Qik.
\newblock \url{http://qik.com/}, 2010.

\bibitem{Song02}
D.~Song, A.~F. van~der Stappen, and K.~Goldberg.
\newblock Exact and distributed algorithms for collaborative camera control.
\newblock In {\em In The Workshop on Algorithmic Foundations of Robotics},
  pages 167--183, Berlin, Germany, 2002. Springer-Verlag.

\bibitem{speex}
Speex: A free codec for free speech.
\newblock \url{http://www.speex.org}.

\end{thebibliography}
%
%

\end{document}